\documentclass{llncs}

\usepackage{graphicx}
\usepackage{comment,array}
\usepackage{cite}
\usepackage{algorithm}
\usepackage{algorithmic}
\usepackage{amssymb}
\usepackage{amsmath}
\usepackage{bbm}

\newcommand\blfootnote[1]{%
  \begingroup
  \renewcommand\thefootnote{}\footnote{#1}%
  \addtocounter{footnote}{-1}%
  \endgroup
}

\title{An SLA-based Advisor for Placement of\\ HPC Jobs on Hybrid Clouds}

\author{Kiran Mantripragada,
Leonardo P. Tizzei,\\
Alecio P. D. Binotto,
Marco A. S. Netto}
\institute{IBM Research}

\newcommand{\procs}{\mathbb{P}}
\usepackage{enumerate}
\usepackage{subfigure}
\usepackage{multirow}
\usepackage{comment}

\begin{document}

\maketitle

\blfootnote{The final publication is available at link.springer.com - ICSOC'15}


\begin{abstract}

Several scientific and industry applications require High Performance Computing (HPC) resources to process and/or simulate complex models.
Not long ago, companies, research institutes, and universities used to acquire and maintain on-premise computer clusters; but, recently, cloud computing has emerged as an alternative for a subset of HPC applications.
This poses a challenge to end-users, who have to decide where to run their jobs: on local clusters or burst to a remote cloud service provider.
While current research on HPC cloud has focused on comparing performance of on-premise clusters against cloud resources, we build on top of existing efforts and introduce an advisory service to help users make this decision considering the trade-offs of resource costs, performance, and availability on hybrid clouds.
We evaluated our service using a real test-bed with a seismic processing application based on Full Waveform Inversion; a technique used by geophysicists in the oil \& gas industry and earthquake prediction. We also discuss how the advisor can be used for other applications and highlight the main lessons learned constructing this service to reduce costs and turnaround times.

\end{abstract}

\section{Introduction}
\label{sec:introduction}

Cloud computing was initially created to host web applications, but has since become an alternative platform for several complex applications, like in Big Data and High Performance Computing (HPC) areas. Typically, these applications execute on on-premise clusters due to their heavy infrastructure requirements; however, as virtualization and network technologies evolve, users can burst to the cloud part or all their workloads to reduce costs and response time.

For cloud environments, one of the key issues is deciding which jobs should be submitted to on-premise resources and which should be burst to the cloud. This decision depends on several variables, such as job execution time on both platforms (local and remote), financial costs, job queuing waiting time, provisioning time in the cloud, among others. End-users need to consider all these parameters to make proper decisions, which increase the complexity for (hybrid) cloud deployments and limit cloud applicability for HPC-based applications, especially when Service Level Agreements (SLAs) are in place. Apart from the number of variables, end-users have to handle their volatility, which increases the chance of wrong decisions, wasting money, or missing important deadlines.

Current work on HPC cloud has mainly focused on understanding the cost-benefits of using cloud over on-premise clusters \cite{ostermann2010performance,napper2009can,vecchiola2009highperformance,gupta2013thewho}; and there is still a gap between understanding the cloud performance and costs, and helping users make decisions on bursting their applications to the cloud. While applications may have overhead in the cloud, mainly due to network performance, cloud is more effective when we consider resource availability---users do not have to wait long time periods in job queues of cluster management systems.

This paper introduces an advisory service to support users in deciding how to distribute computational jobs between on-premise and cloud resources. In summary, our main contributions are:
\vspace{-3mm}
\begin{itemize}
\item An advisory service for bursting applications to the cloud, considering performance and cost difference between cloud and on-premise resources, as well as deadline, local job queue, and application characteristics ($\S$ \ref{sec:method});

\item A case study that shows the advisory service being used by a real world application---a seismic processing technique from the oil \& gas industry. We employed a real test-bed with cloud and on-premise cluster resources, and measured the impact of unreliable execution time predictions on cloud bursting decisions ($\S$ \ref{sec:case-problem}, $\S$\ref{sec:evaluation}).
\end{itemize}

\section{Advisory Service and Policies}
\label{sec:method}

The advisory service considers a user deadline, incurred costs, the on-premise job queue length (local), the provisioning time (cloud), the price ratio between local and cloud for the resource allocation, the type of available hardware, and the estimated execution time for both environments with different configurations. A few of these variables, mainly related to performance and cloud resource setup times, can be specified in SLAs placed between the parties, including users, advisory service, and cloud provider.

The main input parameters to this advisor are the \textit{application profiles} and the \textit{cost models}. The \textit{application profiler} generates a comprehensive application behavior considering infrastructure, financial costs, performance, and number of required processors. Several approaches exist to produce application profiles \cite{calheiros2013emusim,zheng2005simulation,jarvis2006performance,sadjadi2008modeling,yang2005cross}---depending on the application and technique, different accuracy levels can be reached.

The estimation of a \textit{cost model} for a cloud infrastructure is not a difficult task as it can be defined through simple simulations in the cloud provider.
On the other hand, the \textit{cost model} of on-premise resources is challenging since each owner operates financially different, and so we tackle this problem by assuming that \textit{cost model} of on-premise infrastructure is proportional to the \textit{cost model} of cloud infrastructure \cite{marathe2013comparative}.

This section presents the architecture of the advisory service and two policies for HPC job mapping in hybrid clouds. In Section \ref{sec:case-problem}, we instantiate the advisor for a specific use case to show the practical applicability of the proposed service.

\subsection{Architecture}
\label{sec:swarch}

\begin{figure}[!t]
	\centering
	\includegraphics[width=0.99\columnwidth]{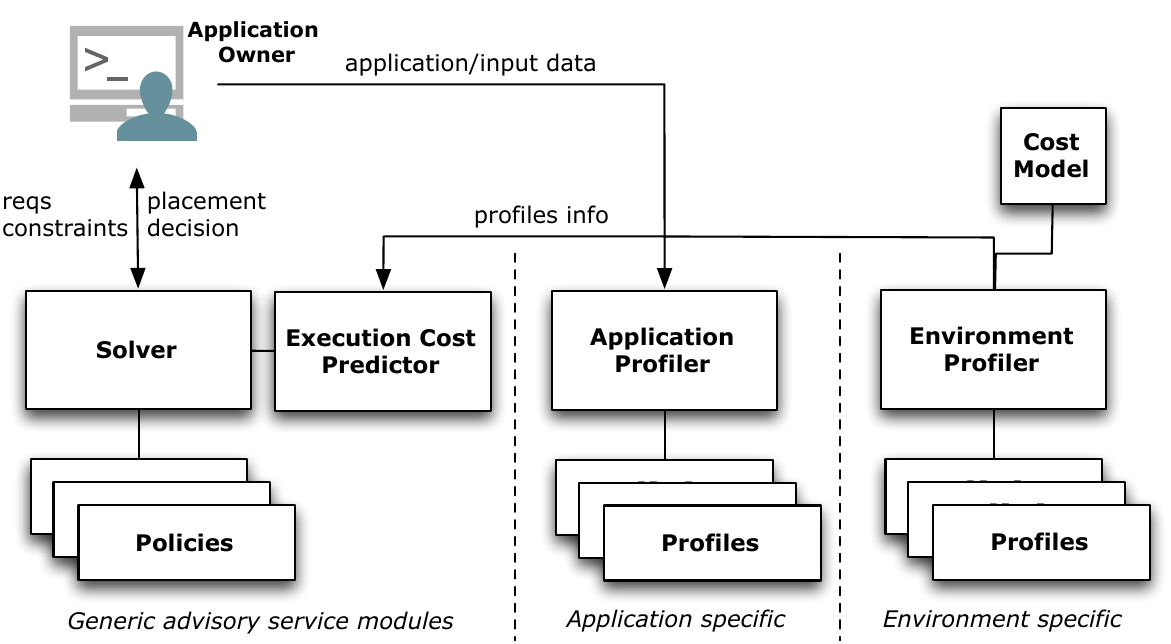}
		\vspace{-3mm}
	\caption{Advisory service architecture}
	\vspace{-5mm}
	\label{fig:swarch}
\end{figure}

The design of the advisory service is intended to be as generic as possible, relying on domain and application specific modules discussed in Section \ref{sec:case-problem}. Figure~\ref{fig:swarch} presents the advisory service architecture containing three groups of modules: generic advisory service modules, application, and environment specific profilers.

\texttt{Solver} receives requirements and constraints to be fulfilled (\textit{e.g.}, cost, execution time, budget, deadline). It calls the \texttt{Execution Cost Predictor}, which estimates the cost of the executions based on their application profile, on the environment cost model, and optionally on execution logs.
The application profile describes the relation between computational resources and execution time, which is important to estimate the total execution cost based on a particular model. Whenever users run applications, details of their executions are stored in a database, similar to the approach described by Binotto \textit{et al.} \cite{binotto2013smartconfig}.
These logs refine future comparisons as they provide historical data that can be analyzed to improve accuracy of the estimations.

Particularly, the \texttt{Solver} module can use the results of the \texttt{Execution Cost Predictor} module to find the best configuration given user's variables and constraints. It first classifies all configurations in two categories: those that fit and those that do not fit the constraints. The earlier are sorted based on the optimization criterion, \textit{i.e.}, time or budget. The solver relies on placement policies---two examples of policies are described in the following section as they are generic enough to be used by several applications and environments.

\texttt{Application Profiler} characterizes an application considering computational resources (\textit{e.g.}, processors, memory, network, I/O) and execution time. In order to compare cloud and on-premise cluster, users should generate application profiles for both environments, which are captured by the \texttt{Environment Profiler} containing the environment \texttt{Cost Model} (Section \ref{sec:cost-model}).

The proposed advisory service currently supports two policies: (i) the maximum \textit{budget} for running jobs, when  users are more concerned about costs; (ii) the maximum \textit{execution time}, when users must meet a deadline to deliver results, and budget is a secondary concern.


\subsection{Budget-aware policy}
\label{sec:budget-aware}
	
For the Budget-aware policy, the user provides a budget restriction and the policy finds the best placement for the job, considering queue length for local cluster, a time for setup and provisioning in the cloud environment, and the price ratio between local and cluster. Here are the steps of this policy:
		
		\begin{enumerate}[(I)]	
		\item Get the \textbf{execution time} for given budget (inverse of Eq. \ref{eq:cost-time-function}, Section \ref{sec:coupled-model});
			
		\item Find the \textbf{number of processors} to be allocated in a \textbf{cloud infrastructure} to meet the execution time  requirement (Eq. \ref{eq:powerlaw}, Section \ref{sec:app-profile});
		
		\item Define a vector with the \textbf{processors per node} distribution based on the available on-premise hardware and the cloud provider. In our experiments, we used the following configuration:
			
			$\procs_{cloud}=\{1, 2, 4, 8, 12, 16\}$,
			$\procs_{local}=\{1, 2, 3, ..., 200\}$.
																				
			Thus, if the advisor computes a number of processors $NProcs_{cloud} = 45$, the distribution will be:
			$ProcPerNode_{cloud}= [16, 16, 13]$. Note that the last node would have $13$ processors, but there may be no such configuration available in the cloud provider. It would be possible to define a distribution of $45$ processors in the following way: $ProcPerNode_{cloud}= [16, 16, 12, 1]$, but it would require an additional processor, which would increase the costs. Instead, the distribution is defined as described in the next step.

		\item Adjust the number of processors based on the available configuration:
		
			New $ProcPerNode_{cloud} = [16, 16, 12]$.
							
			The last node was downsized to $12$ because this is the next available value below $16$ in $\procs_{cloud}$. Since the user is looking for a turnaround time given the budget restrictions (budget-aware policy), the advisor must reduce the number processors trying to meet that budget baseline.

		\item Compute the \textbf{new execution time} and \textbf{turnaround time} from number of processors, queue length, and setup time. The execution time is estimated from equations that define the \textit{Application Profile} (Section \ref{sec:app-profile}).

		\item Compute the costs for the estimated infrastructure using the Cost Model (Section \ref{sec:cost-model}).
															
	\end{enumerate}
At the end, the advisor provides an achievable deadline based on budget restrictions.
The decision would be to place the job in an infrastructure that delivers shorter turnaround time if they are still within the restricted budget.

		
\subsection{Deadline-aware policy}
\label{sec:deadlinea-aware}

This policy determines the required number of processors that meets the deadline in cloud and local environments. The policy steps are:	
		
		\begin{enumerate}[(I)]	
		\setlength\itemsep{-0.8em}
		\item Get \textbf{execution time} for a deadline, setup time and queue length:\\
		 $ t = T - Q $,
		 where $t$ is the execution time, $T$ is the total time (deadline), and $Q$ is the queue length (local) or the setup time (cloud).
		\\
			
		\item Find the \textbf{number of processors} that can fulfill the execution time requirement (Eq. \ref{eq:powerlaw} from the \textit{Application profile});
		\\

		\item Define a vector with the \textbf{processors per node} distribution based on the available hardware (local) and the offers from the cloud provider. In our experiments, we used the following configuration:
			
			$procs\_{cloud}=\{1, 2, 4, 8, 12, 16\}$,
			$procs\_{local}=\{1, 2, 3, ..., 200\}$									

		Thus, if the advisor computes a number of processors $NProcs_{cloud} = 41$, the distribution will be:
			
			$ProcPerNode\_{cloud}= [16, 16, 9]$. Note that, again, the last node would have $9$ processors, but there is no such configuration in the cloud provider. It would be possible to use the following distribution: $ProcPerNode\_{cloud}= [16, 16, 8, 1]$, but it would require an additional processor, which could increase the network communication and possibly the execution time. Instead, we define the distribution of processors as described in the next step.
			\\
			
		\item Adjust the number of processors based on the available configuration:	
			
			New $ProcPerNode_{cloud} = [16, 16, 12]$, where the last node was upscaled to $12$ since this is the next available value greater than $9$ in $\procs_{cloud}$. The same adjustment is done for the \textit{procs per node} in the local HPC. Since users are looking for resources that meet their time constraints, we increase the number of processors in order to satisfy the restricted execution time.
		\\	
		\item Compute the new number of processors according the previous adjustments (new $NProcs = 16+16+12 = 44$);
        \\
		\item Compute the \textbf{new Execution Time} and \textbf{Turnaround Time} from the adjusted number of processors (Eq. \ref{eq:powerlaw}), the queue length, and setup time;
	    \\		
		\item Compute the costs associated with the defined infrastructure (Eq. \ref{eq:total-costs-from-nProcs}).
															
	\end{enumerate}
		\vspace{-1mm}
At the end of this policy, the advisor provides cost values for local and cloud infrastructures, trying to meet the deadline.
If both environments can deliver the restricted deadline, the decision would be based on the Total Costs.

\section{Application Case Study in Oil \& Gas Industry}
\label{sec:case-problem}	

In this section, we present a case study based on Full Waveform Inversion (FWI) application~\cite{Virieux:2009:OFW}. To instantiate the advisory service for this application, we created the application profile and cost model components for SoftLayer cloud provider and an on-premise cluster. Rather than showing an overview of the advisor for several applications, we focused on a detailed explanation of a use case placed into practice. The FWI application contains components of several HPC applications such as communication among multiple processes, solvers for linear systems, matrix operations, among others. There are several ways to create the application profile, which can be implemented and plugged-in into the advisory service \cite{calheiros2013emusim,zheng2005simulation,jarvis2006performance,sadjadi2008modeling,yang2005cross}. The more complex the application behavior, the less accuracy and more time will be required to create the profile. We evaluate the impact of accuracy in Section \ref{sub:evaluation2}.

\begin{figure}[!t]
\centering
\begin{subfigure} 
  \centering
  \includegraphics[width=.65\linewidth]{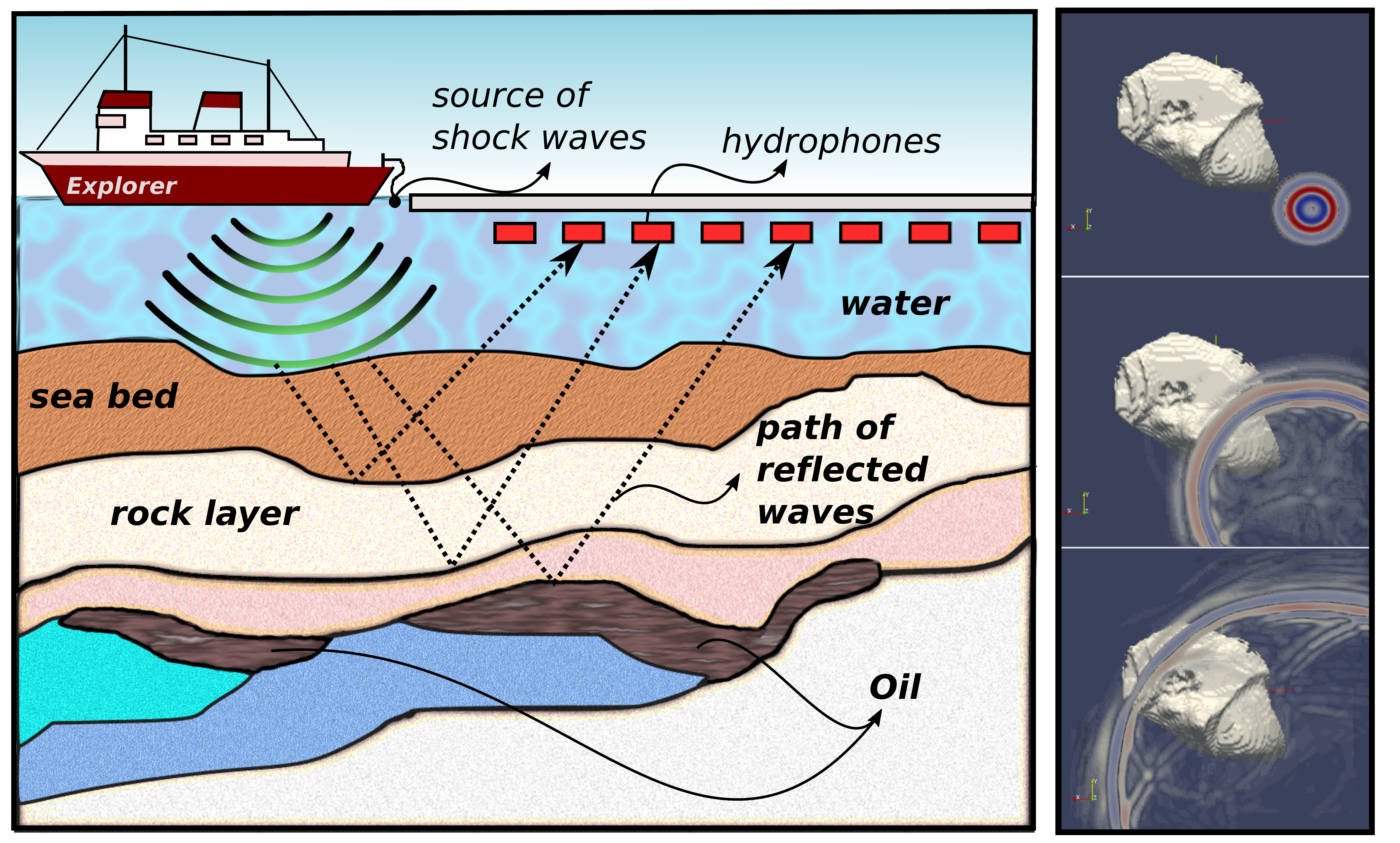}
\end{subfigure}
\caption{FWI data collection and remote visualization}
\label{fig:fwi}
\vspace{-4mm}
\end{figure}

FWI is a numerically challenging technique based on full wavefield modeling of a geological domain that extracts relevant quantitative information from seismograms.
When dealing with realistic physics-based elastic partial differential equations formulation and accurate discretization techniques, such as high order Spectral Element Method, the forward modeling becomes particularly challenging from the computational point of view.
Millions of shots in different positions are calculated over the same data domain and each shot solution evolves along time steps.
At the end, the generated images are composed to produce the final outcome (Figure \ref{fig:fwi}).

			
\subsection{Application Profile}
\label{sec:app-profile}
			
We considered the FWI application, which is based on a multi-resolution, Partial Differential Equation solver to run our experiments. We assume that the profile of such applications can be represented by a power-law function due to its inherent scale-invariance characteristics. For instance, similar applications can be scaled to a finer or coarser grid resolution and will behave similarly, by simple tuning, the coefficients of the power-law function \cite{li2009workload,lowen2005fractal}, for arbitrary constants $a$ and $b$.  From experiments, we observed that the \textit{Application Profile} function can be modeled as:
			\vspace{-1mm}
			\begin{equation}
				t = a \; P^b
				\label{eq:powerlaw}
			\end{equation}
				where $t$ is the execution time, $P$ is the number of processors, and the coefficients $a$ and $b$ are empirically determined. More details on these coefficients can be found in Section \ref{sec:enviroment}, Figure \ref{fig:cost-profile}(b), and Equation \ref{eq:perfcore}. In addition to Equation \ref{eq:powerlaw}, we need to determine the number of processors for a time restriction $t$ as the input parameter. This is solved by inverting Equation \ref{eq:powerlaw}.
										
			
\subsection{Environment Profile: Cost Model}
\label{sec:cost-model}

In order to develop a \textit{cost model}, we collected prices charged by cloud providers. We used the SoftLayer\footnote{SoftLayer website: \url{http://www.softlayer.com/}} cloud infrastructure---similar findings from our experiments could be obtained using other cloud providers as they mostly rely on similar prices and charging models (hourly-based). We consider here homogeneous resources and leave heterogeneity \cite{delimitrou2013qos} as future work. Table \ref{tab:sl-costs} shows SoftLayer fees in US Dollars for different memory size configurations.

			\setlength{\tabcolsep}{1pt}
			\renewcommand{\arraystretch}{1.15}
			\begin{table}[htbp]
				\vspace{-4mm}
				\caption{Costs/hour over number of cores and three memory configurations}
				\vspace{-6mm}				
				\label{tab:sl-costs}
				\scriptsize
				\begin{center}
					\begin{tabular}{rrrr}
					\hline
					\hline
					\multicolumn{1}{l}{} & \multicolumn{1}{c}{\textbf{1GB/proc}} & \multicolumn{1}{c}{\textbf{2GB/proc}} & \multicolumn{1}{c}{\textbf{4GB/proc}} \\ \hline
					\multicolumn{1}{c}{\textbf{N Cores}} & \multicolumn{1}{c}{Cost(\$/hour)} &
					\multicolumn{1}{c}{Cost(\$/hour)} &
					 \multicolumn{1}{c}{Cost(\$/hour)} \\ \hline
					\textbf{1} & 0.040 & 0.059 & 0.098 \\
					\textbf{2} & 0.079 & 0.118 & 0.183 \\
					\textbf{4} & 0.159 & 0.224 & 0.334 \\
					\textbf{8} & 0.306 & 0.416 & 0.591 \\
					\textbf{12} & 0.444 & 0.674 & 0.806 \\
					\textbf{16} & 0.581 & 0.756 & 1.019 \\ \hline
					\end{tabular}
					\vspace{-2mm}
				\end{center}				
			\end{table}

\begin{figure*}[!htb]
	\centering
	\includegraphics[width=0.33\linewidth]{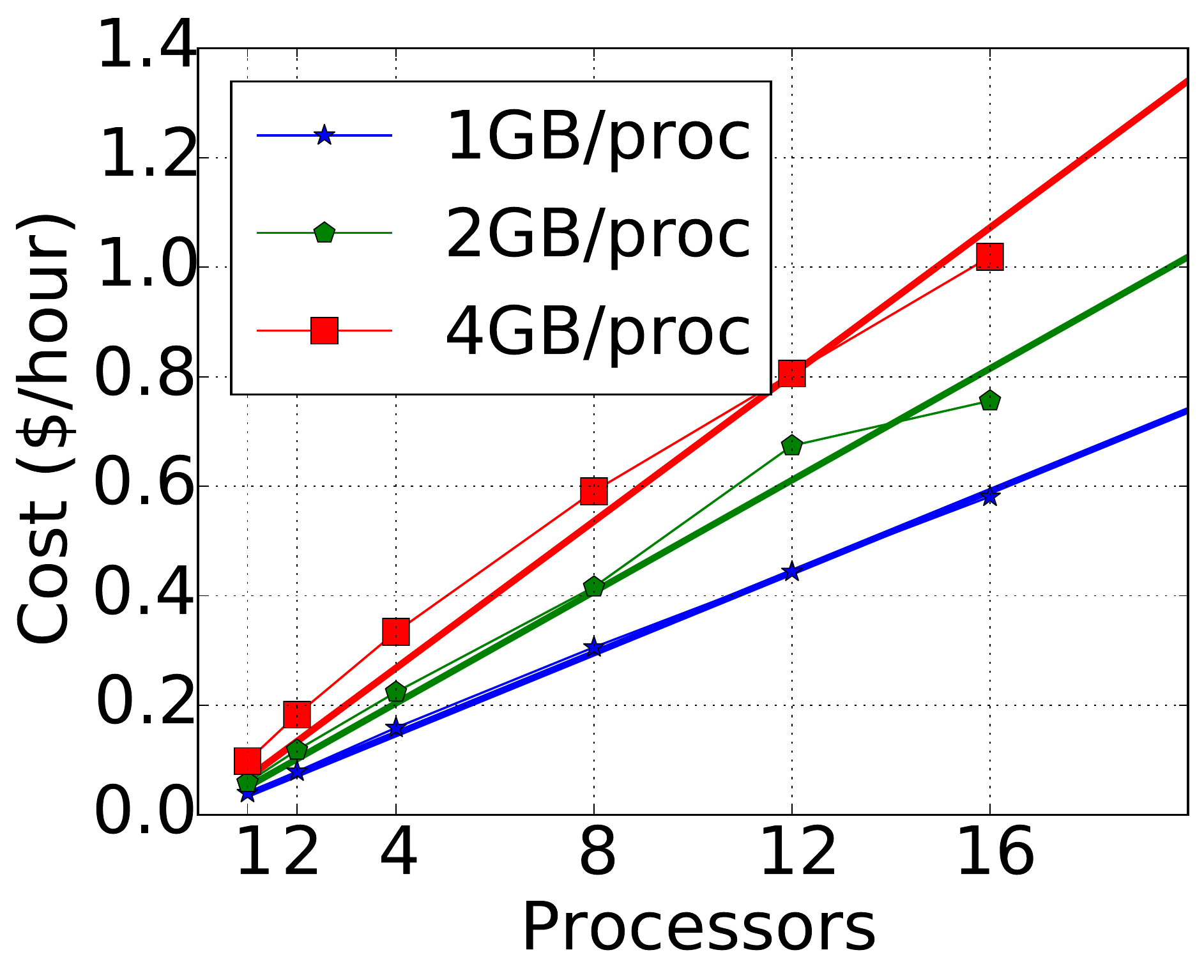}
	\hfill
	\includegraphics[width=0.32\linewidth]{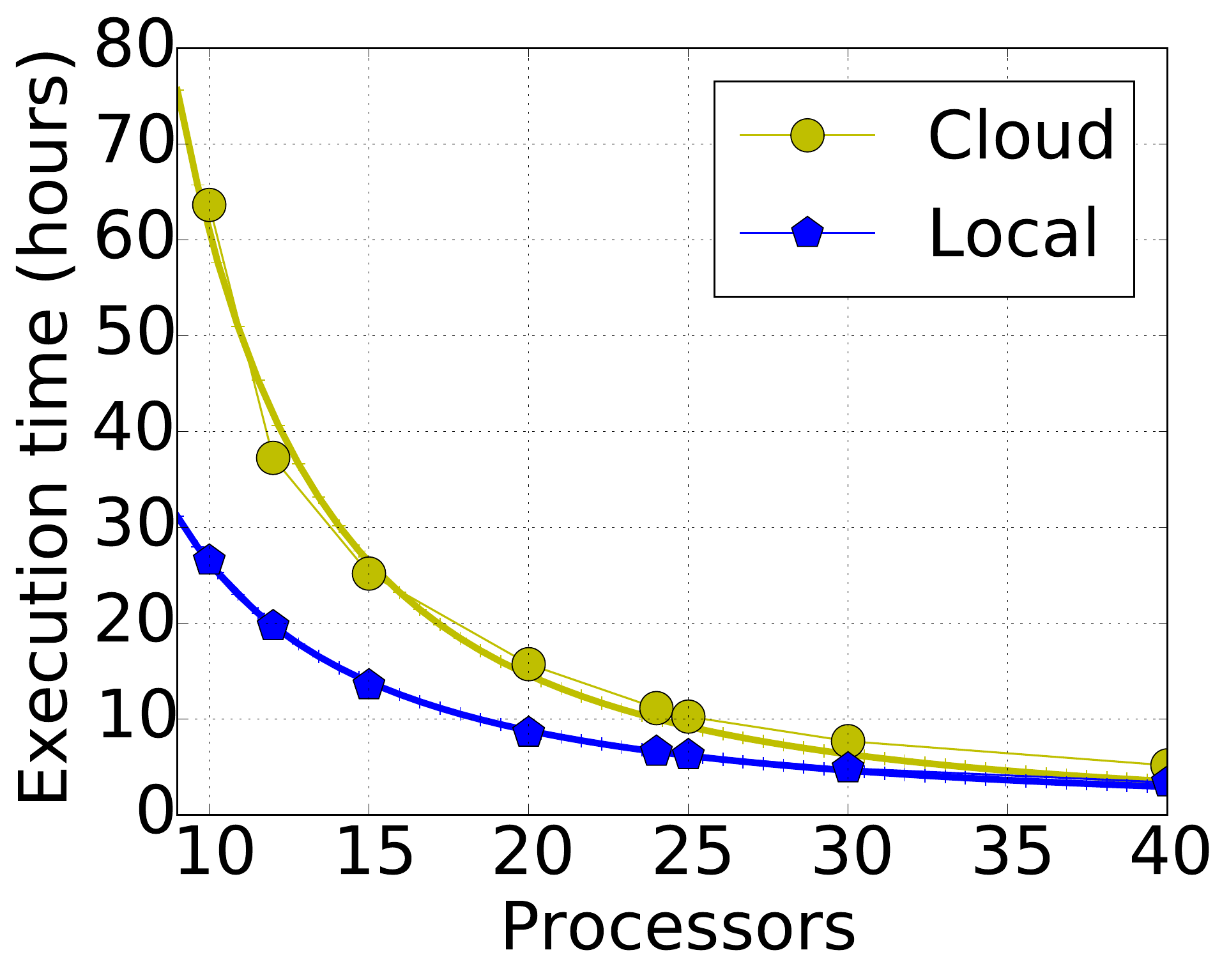}	
	\hfill
	\includegraphics[width=0.33\linewidth]{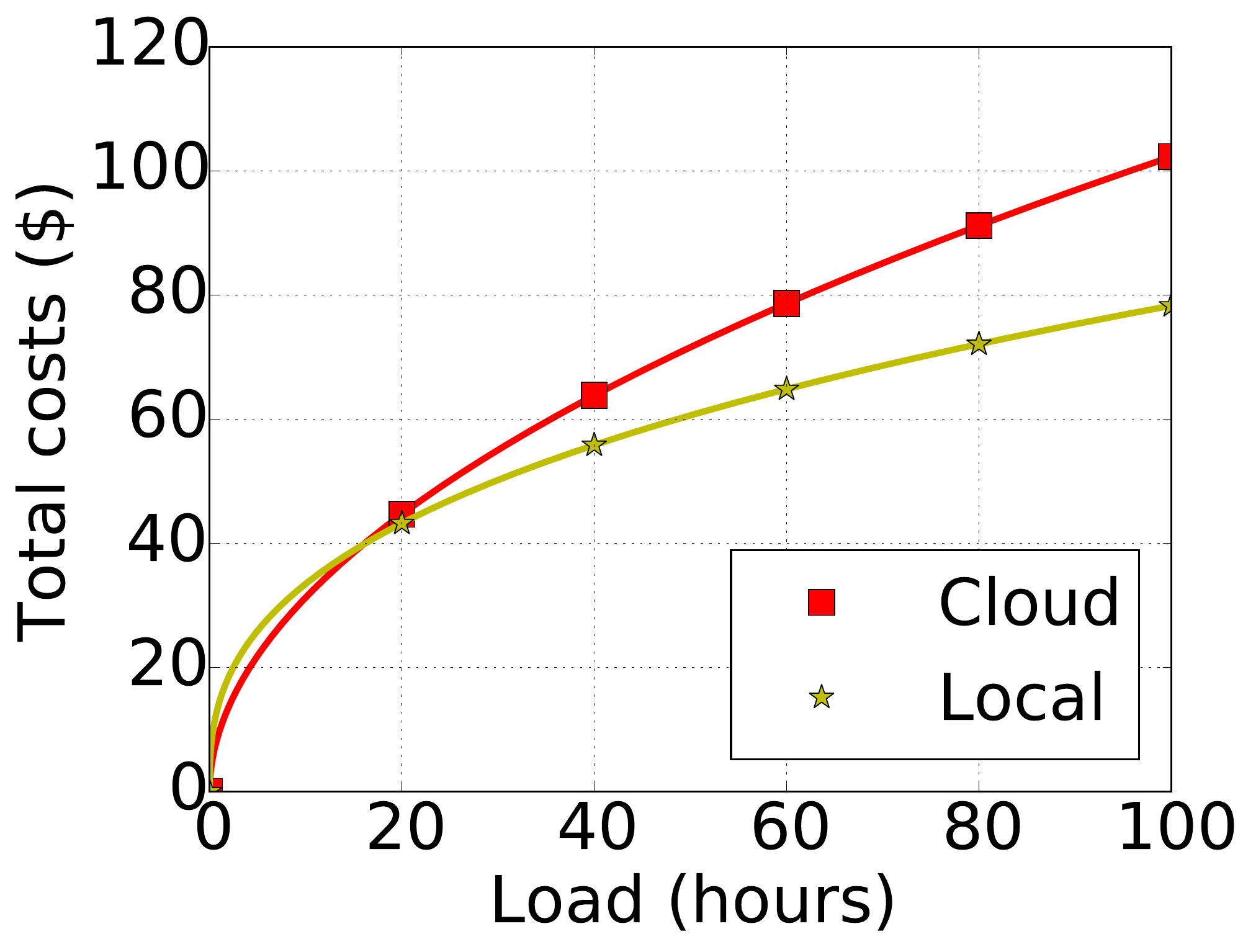}		
	\caption{Cost Model: (a) Listed price per node (cloud only); (b) Workload profile; (c) prices for 4GB RAM/proc and price ratio k=1}
	\label{fig:cost-profile}
	\vspace{-5mm}
\end{figure*}

We observed that the hourly-rate for provisioning nodes is a linear relationship to the number of processors \textbf{$P$}.
These functions are shown in Figure \ref{fig:cost-profile}(a) and were fitted from real prices offered in SoftLayer (Table \ref{tab:sl-costs}).
The ratio ``price per hour'' can be described by a general linear equation $(C_h = \alpha P + \beta)$.
	For simplification purposes, we assume the offset coefficient $(\beta)$ can be neglected and the ``price per hour'' is directly proportional to the number of processors. We also claim that $C_h$ can be written as $dC/dT$, by considering a continuous execution time:
	\begin{equation}
		\frac{dC}{dt} = \alpha P
		\hspace{20px}
		\Rightarrow
		\hspace{20px}
		dC = (\alpha P) dt 		
		\label{eq:price-hour}
	\end{equation}
	where $dC/dt$ is the hourly-rate for nodes and $\alpha$ is a linear coefficient determined empirically.
			
In order to quantify the total cost for a given turnaround time, we can integrate Equation \ref{eq:price-hour} over time (remembering that the number of processors $P$ does not change with time):

	\begin{equation}
		C = \int_0^T (\alpha P) dt
		\hspace{20px}
		\Rightarrow
		\hspace{20px}				
		C = T \, (\alpha P)
		\label{eq:total-costs-from-nProcs}				
	\end{equation}

	where $T$ is the turnaround time and $C$ is the total cost for a given number of processors $P$ and turnaround time $T$.
	
The integration constant is $zero$, since $C(0)=0$, meaning there is no costs when execution time is also $zero$.

It is not a trivial task to model costs of on-premise HPC clusters \cite{marathe2013comparative}, as it is dependent on acquisition time, depreciation, occupancy, and maintenance costs. Therefore, in order to simplify the on-premise cost model, we assume it is proportional to cloud costs \cite{gupta2013thewho,marathe2013comparative}:
	\begin{equation}
		\left(\frac{dC}{dt}\right)_{local} = K \,* \, \left(\frac{dC}{dt}\right)_{cloud}
		\label{eq:price-ratio-K}
	\end{equation}
	where $(dC/dt)_{local}$ and $(dC/dt)_{cloud}$ \,\, are hourly prices for local and cloud infrastructures respectively, and $K$ is the local-cloud price ratio.	
	All remaining equations can be derived from Equation \ref{eq:price-ratio-K} in a similar way done for cloud costs (Equations \ref{eq:price-hour} to \ref{eq:total-costs-from-nProcs})
	\vspace{-1mm}
	\begin{equation}
		C = T \, (K \, \alpha \, P)
\end{equation}
	

\subsection{Integrating application profile and costs model}
\label{sec:coupled-model}
		In order to find the total costs for a given turnaround time, we integrated this function over time.
		From Equations \ref{eq:price-hour} and \ref{eq:powerlaw}, the following can be written:
		\vspace{-1mm}
		\begin{equation}
			dC =    \left[  \alpha \left(\dfrac{t}{a}\right)^{\left(\frac{1}{b}\right)}  \right] \, dt  \nonumber			
			\hspace{10px}
			\Rightarrow
			\hspace{10px}		
			\int dC = \int_0^T \left[ \alpha \left(\dfrac{t}{a}\right)^{\left(\frac{1}{b}\right)}  \right] \, dt \nonumber
		\end{equation}
	
	 We have $C$, which is the total cost for a given time $T$, a cost model (from $\alpha$), and the application profile (from $a$ and $b$):
		\begin{equation}
			C = a \, \alpha \left[   \frac{\left(\frac{T}{a}\right)^{(1+\frac{1}{b})}}{1+\frac{1}{b}}   \right]
			\label{eq:cost-time-function}
		\end{equation}

Equation \ref{eq:cost-time-function} defines the coupled \textit{``Profile-Cost model''}, and are used in the ``Budget-aware policy".
Figure \ref{fig:cost-profile}(c) shows the costs function $C(T)$ for a load (execution time X number of processors).

The inverse function $C^{-1} = T(C)$ is also useful when we have a budget restriction and want to look for a total execution time.

\section{Evaluation}
\label{sec:evaluation}	

The goals of the evaluation are to understand: (i) the financial and time saving benefits of the advisor (Section~\ref{sub:evaluation1}); and (ii) how the advisor is dependent from the accuracy of the application profile (Section~\ref{sub:evaluation2}). We compared the advisor against four policies:

\begin{itemize}
 \item \textbf{Always-local:} chooses to always submit jobs to the on-premise environment---it represents users who do not want or cannot move their jobs to the cloud. We used this policy as baseline for comparison because it still represents the most conservative and traditional behavior of HPC users;
 \item \textbf{Always-cloud:} chooses to always submit jobs to the cloud---it represents users who do not have access to an on-premise cluster or are willing to test the cloud to avoid acquiring a new cluster in the future;

 \item \textbf{Random:} randomly decides between cloud and local environments---it is an attempt to represent users who do not have any supporting mechanism or intuition to know where to run their jobs;
 \item \textbf{Worst-case:} chooses the opposite environment provided by the advisor---it represents hypothetical users who make extremely wrong decisions. This helps us understand how much a user can loose with such decisions.
\end{itemize}

This is not an exhaustive list of policies that we could use for comparison. It represents a few extreme cases to understand how good or bad resource allocation decisions can impact costs and turnaround times if no advising mechanisms are provided to the user. Several other policies \cite{assuncao2009evaluating,unuvar2014hybrid} could be leveraged or further developed in comparison to those used by the advisor. Finding the optimal resource allocation is out of the scope of this paper.

To compare these decision policies with the advisor, we executed both budget-aware and deadline-aware policies varying the following input data, which can be part of SLAs: deadline ranges from 1 to 100 hours; budget ranges from 10 to 100 USD; queue size time ranges from 1\% to 50\% of the deadline; setup time ranges from 1\% to 50\% of the deadline; price ratio between cloud and local environments, with the price of cloud environment ranging from 70\% to 340\% the price of the local environment; total of 28,000 executions per policy.

All ranges have been defined to cover a wide spectrum of possible inputs.
The ranges for the budget, deadline, and setup time are based on our experience with the FWI application.
The price ratio is based on HPC cloud literature \cite{gupta2013thewho}.
The FWI case study profile was generated using a single input data set, which described the size of the domain, the precision of the output image, and the varying number of processors from 10 to 40.


The advisor computes the costs and turnaround time for both environments for each set of input variables, thus executing both deadline-aware and budget-aware policies.
For instance, if the deadline-aware policy is executed, it computes the costs for both environments, say $Cost_{cloud}$ and $Cost_{local}$, in order to choose the cheapest one.
For each result, the advisor calculates the relative difference between the costs of both environments in the following way:

\begin{equation}\label{eq:relativeCost}
\frac{min(Cost_{cloud},Cost_{local})-Cost_{local}}{Cost_{local}}
\end{equation}
where $Cost_{local}>0$. When the budget-aware policy is executed, the advisor calculates the relative difference of the turnaround time in a similar manner.
The results of the other decision policies used for comparison are also relative to the always-local decision policy.

\subsection{Environment Setup}
\label{sec:enviroment}

We selected two representative environments, described in Table~\ref{tab:environments}, to compare the target application on cloud and on-premise environments.

\begin{table}[!htb]
	\vspace{-3mm}
	\caption{Cluster and SoftLayer details}
	\vspace{-6mm}
	\label{tab:environments}
	\begin{center}
	\scriptsize
	\begin{tabular}{ccc}\hline\hline
		          & \textbf{Cluster} & \textbf{SoftLayer} \\ \hline
		processor (GHz) & 2.80 & 2.60 \\
		cores per processor & 10 & 4\\
		cache size (KB) & 25600  & 20480 \\
		total memory (MB) & 132128 & 65712 \\
		network & ethernet/infiniband & ethernet\\
		operating system & RHEL & CentOS\\ \hline
	\end{tabular}
	\vspace{-5mm}
	\end{center}
\end{table}

The FWI application contains a set of parameters such as number of degrees of freedom, polynomial order, data domain, among others, and it is a highly CPU and memory intensive. Details of the application parameters are not relevant to the experiments presented here (they are addressed elsewhere \cite{Virieux:2009:OFW}).

Application profiles provide information about the behavior of the application in terms of metrics, such as total elapsed time, number of cores, memory, nodes, network, and costs. The selection of these metrics depends on the constraints and variables that end-users want to optimize. For instance, Figure~\ref{fig:cost-profile}(b) shows how the total elapsed time for executing application changes as the number of processors increases in the two distinct environments.

We derived the power-law scaling function, as described in Section \ref{sec:app-profile}, from the experiments (Equation \ref{eq:powerlaw}). The coefficients $a$ and $b$ were computed through non-linear least squares curve fitting. Thus, we have:

\begin{equation}
	t_{local} =  1013.50 \, P_{local}^{-1.58}
	\hspace{30px}
	and
	\hspace{30px}
	t_{cloud} =  7004.86 \, P_{cloud}^{-2.06}	
	\label{eq:perfcore}
\end{equation}

\begin{figure*}[!b]
	\centering
	\includegraphics[width=1\linewidth]{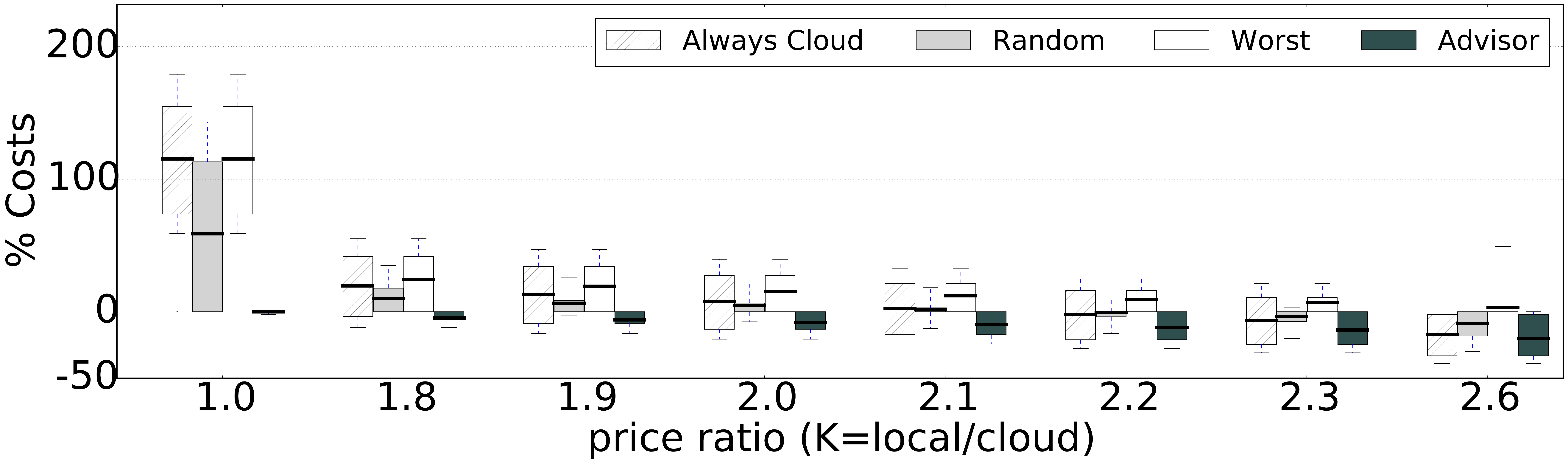}
	\vspace{-8mm}
	\caption{Deadline-aware policy: the lower the costs the better the policy}
	\label{fig:subDeadline}
	\vspace{5mm}	
	\includegraphics[width=1\linewidth]{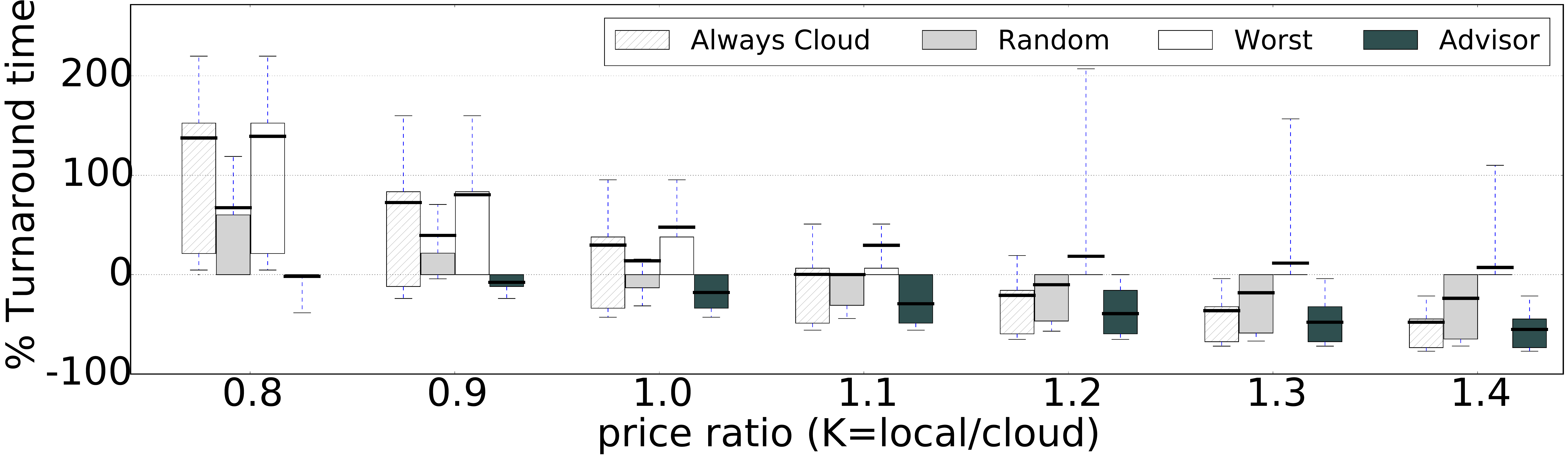}
	\vspace{-8mm}	
	\caption{Budget-aware policy: the lower the time, the better the policy}
	\label{fig:subBudget}
\end{figure*}


\subsection{Results: Costs and Time Savings}
\label{sub:evaluation1}

The results of the comparison are presented in Figures~\ref{fig:subDeadline} and \ref{fig:subBudget}.
The $y$ axis of Figure~\ref{fig:subDeadline} represents the relative cost as calculated by Equation~\ref{eq:relativeCost}, while the $y$ axis of Figure~\ref{fig:subBudget} describes the relative turnaround time.
In both figures, $y$ axes are in a symmetrical log scale and the $x$ axes depict the price ratio.
The higher the price ratio, the higher the costs for local environment compared to cloud.
In both figures, results were stabilized outside the plotted ratios, and therefore we omitted them.

Figure~\ref{fig:subDeadline} shows the results for the deadline-aware policy. When local environment is cheaper ($K<$ 1) and even 80\% ($K$=1.8) more expensive than cloud environment, it delivers the cheapest cost most of the time.
For instance, when the price ratio is 1.8, local environment provides the cheapest cost around 71\% of the instances.
The reason for this is that local environment showed the best computing performance for executing the target application, as depicted in Figure~\ref{fig:cost-profile}.
Thus, even when the local environment is around 80\% more expensive than cloud, its performance counterbalances its cost.
When the price ratio reaches 2.2, the local environment is surpassed by the cloud in terms of cost, \textit{i.e.}, cloud is the cheapest environment around 56\% of the time and this percentage becomes greater as the price ratio grows, as expected.

Figure~\ref{fig:subBudget} depicts the results for the budget-aware policy.
Similarly to the deadline-aware policy, the always-cloud policy is comparable to worst-case and advisor is comparable to always-local when the local price is equal to or below the cloud price.
However, for higher price ratios, always-cloud policy surpasses always-local faster than deadline-aware policy.
The turning point occurs when price ratio is around 1.0: always-local is on average 30\% faster than always-cloud, but the median is -0.12; that is, always-cloud is half the time at least 12\% faster than always-local.
Although the local environment has better computing performance results over cloud, when they have a similar price (\textit{i.e.}, price ratio around 1.0), the decision on where to run for a given budget is not obvious due to the other input parameters (queue length and setup time) affecting the turnaround time.

When the advisor calculates an execution time to meet the budget, the coupled model yields a solution that is near-optimal in terms of execution time. We observed that searching for the optimal execution time is not worthwhile since the adjustments over the infrastructure to meet the available configurations overpass intermediate values found in the optimal solution. For instance, an estimated number of processors $NProcs=9.5$ must be adjusted to $10$ or $9$, according to the policies.

Both figures show that the advisor supports the selection of the environment that best suits users' needs.
The lack of such supporting tool might cause unnecessary costs or waste of time as it is illustrated by the worst-case decision policy.
Furthermore, some of the input variables (\textit{e.g.}, queue size time) can change very frequently thus making impossible to manually calculate the best environment to run an application.


\subsection{Results: Accuracy of the Application Profile}
\label{sub:evaluation2}

We defined the range of inaccuracies from $-90$\% (\textit{i.e.}, $-0.9$) error to $100$\% (\textit{i.e.}, $1.0$) error, aiming to cover a wide spectrum of such profiles.
For each set of input data, the application profile specifies the amount of infrastructure necessary to meet deadline or budget constraints depending on the policy.
Let us say this infrastructure has $100$ cores disregarding the policy; after injecting the error within the aforementioned range, it will have from $10$ (i.e., $100*(1.0-0.9))$ to $200$ cores (i.e., $100*(1.0+1.0))$.

For each estimation of the advisor, the input data also varied in the same way that was described in previous evaluation (Section~\ref{sub:evaluation1}), which means that each policy has been executed 28,000 times for each inaccurate profile.
After the execution, the advisor provided either the same decision that was computed using the accurate profile or a different one. Even if the decision was the same, it is usually based on slightly different results.
Thus, we measured whether the decision is the same or not, and the relative difference between results using an inaccurate and the accurate profile.
These relative differences, when executing the deadline-aware policy, were calculated as follows:

\begin{equation}\label{eq:relativeDelta}
\frac{Cost_{Inaccurate}-Cost_{Accurate}}{Cost_{Accurate}}
\end{equation}
where $Cost_{Inaccurate}$ and $Cost_{Accurate}$ are the costs measured using inaccurate and accurate profiles, respectively.
The relative differences were calculated similarly when executing the budget-aware policy.
Table~\ref{tab:error} shows a comparison of the results provided by the advisor using inaccurate and accurate profiles.
The first column compares the decision of the advisor using both profiles: $=$ means same decision and $\neq$ otherwise.
For each policy, this table shows the average of the relative differences (avg), their standard deviation (std) and total number of decisions (size) for each inaccurate profile.
So, for each inaccurate profile, the sum of equal and different decisions is 28,000.

\begin{table}[!htb]
\centering
\vspace{-3mm}
\caption{Results from the advisor using inaccurate and accurate profiles}
\label{tab:error}
\scriptsize
\begin{tabular}{c|c|c|c|c|c|c|c}\hline
 decision & error& \multicolumn{3}{|c|}{deadline-aware}& \multicolumn{3}{|c}{budget-aware}  \\ \cline{3-8}
      &      & avg  & std  & size                  & avg & std & size \\ \hline
  =   & \multirow{2}{*}{-0.9} &-0.8  & 0.3  & 17293 (62\%)                 &-0.2 & 0.7 & 13068 (47\%)\\ \cline{1-1} \cline{3-8}
$\neq$&      	             &-1.0  & 0.0  & 10707 (38\%)          &-0.4 & 0.5 & 14932 (53\%) \\ \hline \hline
  =   & \multirow{2}{*}{-0.5} &-0.4  & 0.3  & 25356 (91\%)          & 0.3 & 0.6 & 23236 (83\%)\\ \cline{1-1} \cline{3-8}
$\neq$&     		     &-0.6  & 0.0  & 2644 (9\%)            & 0.0& 0.4 & 4764 (17\%)\\ \hline \hline
  =   & \multirow{2}{*}{-0.1} &-0.1  & 0.1  & 26004 (93\%)                 & 0.1& 0.4 & 27272 (96\%)\\ \cline{1-1} \cline{3-8}
$\neq$&                       & 0.1  & 0.3  &  1996  (7\%)           & 0.2& 0.3 & 728 (4\%)\\ \hline \hline
  =   &  \multirow{2}{*}{0.1} & 0.1  & 0.1  &27115 (97\%)            & 0.1& 0.1 &27829 (99\%) \\ \cline{1-1} \cline{3-8}
$\neq$&  		     & 0.1  & 0.1  & 885 (3\%)             & 0.1& 0.0 &171 (1\%) \\ \hline \hline
  =   &  \multirow{2}{*}{0.5} & 0.4  & 0.4  &26597 (95\%)            & 0.1& 0.3 &26685 (95\%) \\ \cline{1-1} \cline{3-8}
$\neq$&                       & 0.8  & 0.2  & 1403 (5\%)             & 0.2& 0.2 &1315 (5\%) \\ \hline \hline
  =   &  \multirow{2}{*}{0.9} & 0.9  & 0.8  &25982 (93\%)                  & 0.0& 0.3 &25807 (92\%) \\ \cline{1-1} \cline{3-8}
$\neq$&                       & 1.5  & 0.4  &2018 (7\%)              & 0.3& 0.3 &2193 (8\%) \\ \hline
\end{tabular}
\end{table}

As the error gets close to zero, the percentage of same decisions increases.
Even when the error is 0.9, the advisor computed the same decision  for deadline-aware policy around 93\% of the times and for budget-aware policy around 92\% of times.
The number of different decisions for the budget-aware is greater than the number of same decisions (53\% against 47\%, respectively) only when the error is -0.9.
For this error, the advisor proposed the same decision for the deadline-aware policy around 62\% of the time.

For most of the inaccuracy ranges, the number of equal decisions is far greater than the number of different decisions.
That is, even if the application profile is inaccurate, it has a minor decision impact. Thus, the advisor provides evidence that it is resilient to inaccuracies in the application profile.
One reason for this is the wide spectrum of data that has been exercised.
In some situations, the difference between choosing cloud or local is so great that the inaccuracy has little to no impact on job(s) placement decision.

These results also show that as inaccuracy gets close to zero, the number of correct decisions increases, as expected.
When the inaccuracy is close to zero, there is a higher chance that the infrastructure calculated by the application profile is not relevant for the final result.
For example, an inaccuracy of -10\% of 100 cores would calculate 90 cores; instead of running in 1 hour and 30 minutes, the job(s) would run 1 hour and 50 minutes.
For a cloud environment that rents VMs per hour, the result would be 2 hours for both options.

When the error is -0.9, the number of different decisions achieved its peak for both policies.
On the other hand, when the error is 0.9, the number of different decisions is small compared to the number of equal decisions.
Despite the absolute values of both errors being the same, the number of different decisions provided by the advisor using these inaccurate profiles is distinct.
The results provided by the advisor are based on application profile and the integrated profile-cost model, which are not linear, as depicted in Figures~\ref{fig:cost-profile}(b) and (c). This explains why these inaccurate profiles provided distinct rates.

\section{Related Work}
\label{sec:rw}	

Scientific and engineering applications usually require HPC platforms and are currently being tested on modern cloud platforms that are now starting to incorporate HPC features.
For instance, UberCloud experiment reports \cite{ubercloud2014,ubercloud2013} contain a set of use cases and challenges when moving HPC applications to the cloud. These challenges include data transfer from and to the cloud, unpredictable costs when using cloud resources which depend on workload demands and algorithm complexity, lack of easy and intuitive self-service administration, long waiting times to start-up VMs (at unacceptable level for some end-users), among others.

Ostermann \textit{et al.} \cite{ostermann2010performance} evaluated whether performance of cloud environments are sufficient for scientific computing. Their results showed that cloud environments were insufficient for scientific computing at the time of their publication.
At the same time, Napper and Bientinesi \cite{napper2009can} used Linpack benchmarks to evaluate whether cloud could potentially be included in the top 500 list of supercomputing. Their results showed that the performance of single cloud nodes was as good as nodes in HPC systems, however, memory and network were not sufficient to scale the application. Vecchiola \textit{et al.} \cite{vecchiola2009highperformance} also evaluated the use of clouds for scientific applications. They concluded that clouds are effective for conducting scientific experiments, but the trade-off between costs and performance has to be evaluated case by case.

Following a new phase of HPC cloud studies, Mateescu \textit{et al.} \cite{mateescu2011hybrid} evaluated a set of benchmarks and complex HPC applications on a range of platforms, both in-house and in the cloud.
The studies showed cloud effectiveness for such applications mainly in the case of complementing supercomputers.
Gupta and Milojicic \cite{gupta2011evaluation} highlighted that cloud can be suitable for some HPC applications, but not all. The same group of authors \cite{gupta2013thewho} performed a set of experiments using benchmarks and complex HPC applications on various platforms, including supercomputers and clouds, to answer the question ``why and who should choose cloud for HPC, for what applications, and how should cloud be used for HPC?''.

More recently, Belgacem and Chopard \cite{belgacem2014hybrid} evaluated a computational fluid dynamics application over a heterogeneous environment of a cluster and cloud.
They ratified a previous work \cite{zaspel2011massively} that demonstrated the potential of clouds to fluid dynamics. Mantripragada \textit{et al.} \cite{Mantripragada2015cloudcluster} proposed a method to use a hybrid approach using a local cluster plus the cloud to dynamically boost computing intensive applications dealing with data partitioning respecting performance deadlines.
The results using a seismic application showed that the approach is feasible in the case of a seamless connection between both environments despite of synchronization concerns. Outside the HPC efforts, Unuvar \textit{et al.} \cite{unuvar2014hybrid} introduced a hybrid cloud placement algorithm, which focus on application structure to allocate multiple VMs.

Although these findings do not propose advisory tools, their empirical studies open an opportunity for proposing new tools focusing on a heterogeneous approach over local and cloud, like ours.


\section{Conclusions}
\label{sec:conclusion}

We introduced an advisory service based on two policies to help users decide where to run their HPC jobs: on-premise \textit{versus} cloud.
It considers the application profile, job waiting queue, costs, budget, and deadline constraints. We used a real testbed and seismic application in the area of oil \& gas industry to perform a comprehensive set of experiments.

The advisory service is composed of modules that can be extended/plugged-in to have more refined \textit{Application Profiles} and to suit other applications. Further investigations will be required to collect data from a wide range of applications. The main lessons from our study are:
(i) in HPC cloud, apart from resource performance, it is important to consider the time a user has to wait in a job queue of the on-premise environment compared to the overhead of cloud resources---more relevant is the total turnaround time, as also pointed by Marathe \textit{et al.} \cite{marathe2013comparative};
(ii) it is possible to consider an advisory service for HPC hybrid clouds even without having highly precise application/job profiles---however, very inaccurate profiles may generate negative impact on costs and turnaround delays;
(iii) the higher the cost differences between cloud and on-premise resources the higher the savings brought by an advisory service for resource selection on hybrid clouds.


\section*{Acknowledgment}
\label{sec:ackwnoledgement}	

We thank Eduardo Rodrigues and Nicole Sultanum for their comments on this paper. This work has been partially supported by FINEP/MCTI under grant no. 03.14.0062.00.

\bibliographystyle{splncs03}
\bibliography{references}
\end{document}